# Pan-European fuel map server: an open-geodata portal for supporting fire risk assessment

Erico Kutchartt [a,b], José Ramón González-Olabarria [a,c], Núria Aquilué [a,c], Jordi Garcia-Gonzalo [a,c], Antoni Trasobares [a], Brigite Botequim [d,e], Marius Hauglin [f], Palaiologos Palaiologou [g], Vassil Vassilev [h], Adrian Cardil [a,i], Miguel Ángel Navarrete [i], Christophe Orazio [j], Francesco Pirotti [b,k*]

[a] Forest Science and Technology Centre of Catalonia (CTFC). Carretera de Sant Llorenç de Morunys, Km 2, 25280 Solsona, Spain

[b] Department of Land, Environment, Agriculture and Forestry (TESAF), University of Padova. Via dell'Università 16, 35020 Legnaro (PD), Italy

[c] Joint Research Unit CTFC – AGROTECNIO. Carretera de Sant Llorenç de Morunys, Km 2, 25280 Solsona, Spain

[d] CoLAB ForestWISE - Collaborative Laboratory for Integrated Forest & Fire Management. Quinta de Prados 5001-801 Vila Real, Portugal

[e] Forest Research Centre, School of Agriculture (ISA), University of Lisbon. Tapada de Ajuda 1349-017 Lisbon, Portugal

[f] Division of Forestry and Forest Resources, Norwegian Institute of Bioeconomy Research (NIBIO). Postboks 115, NO-131 Ås, Norway

[g] Department of Forestry and Natural Environment Management, Agricultural University of Athens (AUA), Karpenisi 36100, Greece

[h] Space Research and Technology Institute, Bulgarian Academy of Sciences, str. "Acad. Georgy Bonchey" bl. 1, 113 Sofia, Bulgaria

[i] Technosylva. Parque Tecnológico de León, 24009 León, Spain

[j] Institut Européen de la Forêt Cultivée (IEFC), 69 Route d'Arcachon, 33610 Cestas, France

[k] Interdepartmental Research Center of Geomatics (CIRGEO), University of Padova. Via dell'Università 16, 35020 Legnaro (PD), Italy

**Abstract**

Canopy fuels and surface fuel models, topographic features and other canopy attributes such as stand height and canopy cover, provide the necessary spatial datasets required by various fire behaviour modelling simulators. This is a technical note reporting on a pan-European fuel map server, highlighting the methods for the production and validation of canopy features, more specifically canopy fuels, and surface fuel models created for the European Union's Horizon 2020 "FIRE-RES" project, as well as other related data derived from earth observation. The aim was to deliver a fuel cartography in a findable, accessible, interoperable and replicable manner as per F.A.I.R. guiding principles for research data stewardship. We discuss the technology behind sharing large raster datasets via web-GIS technologies and highlight advances and novelty of the shared data. Uncertainty maps related to the canopy fuel variables are also available to give users the expected reliability of the data. Users can view, query and download single layers of interest, or download the whole pan-European dataset. All layers are in raster format and co-registered in the same reference system, extent and spatial resolution (100 m). Viewing and downloading is available at all NUTS scales, ranging from country level (NUTS0) to province level (NUTS3), thus facilitating data management and access. The system was implemented using R for part of the processing and Google Earth Engine. The final app is openly available to the public for accessing the data at various scales.

**1. Introduction**

Recently, the analysis and integration of vegetation fuel data have become a critical issue for understanding and preventing the impacts of wildfires since they are an important input for various fire simulators used in Europe, USA, South America and Australia (Cardil et al. 2023). Notable forest fire simulation software includes FARSITE (Finney, 2004), FlamMap (Finney, 2006), Prometheus (Tymstra et al. 2010), Wildfire Analyst (Monedero et al. 2019), and Cell2Fire (Pais et al. 2021), among other alternatives. Accessing to comprehensive geo-spatial data for fuel conditions, expected fire behaviour, and risk assessment is thus of prime necessity, especially when all these layers are combined to inform wildfire management agencies on decisions needed to be taken for prevention and suppression purposes. To address this need, few data servers have been developed in the past at regional and national levels to provide the necessary geospatial data functionality that is capable of predicting potential fire behaviour. Krsnik et al. (2020) have created a data server for Catalonia that helps to evaluate fuel hazards and fire behaviour to mitigate the negative impact of wildfires under different meteorological scenarios. Rollis et al. (2009) developed the LANDFIRE, which provides the current state of vegetation, wildland fuel, and fire regimes across the United States at 30 m raster grids. In addition, GIP ATGeRi in France is maintaining a 30 m resolution fuel map for spatial planning and risk management. The above-mentioned open servers have proved their value in supporting decision-making and preventing and managing forest fires. However, there is an information gap regarding data availability on a continental scale, especially in Europe, where data needs to be harmonized and delivered on a single open access server for large scale landscapes.

Nowadays, a wide range of remote sensing products can be used to assess the terrain and vegetation characteristics using passive and active sensors (Heisig et al. 2022). These new sensors can feed models to spatially map with high accuracy the main fire drivers, identified by three components, namely, topography, weather, and fuels (Keane et al. 2001). Remote sensing derived products at the global scale are already available. Topography variables (elevation, slope, and aspect) were produced by ALOS World 3D (Tadono et al. 2016), canopy features, such as canopy height maps were recently provided at high-resolution by Tolan et al. (2024) using a convolutional network trained on GEDI observations, while canopy cover fraction was generated originally by Hansen et al. (2013) based on Landsat data and then by Liu et al. (2024) using MODIS observations. Topography, canopy cover and canopy height are accessible and available for download at worldwide scale. Other key input variables for fire behaviour modelling and fire risk assessment are less available at continental scale, and sparse data is far to be harmonised through the same methods and spatial resolution. Specifically, two other canopy features (CBH and CBD) and the surface fuel model maps are partially available in Europe and globally (e.g., Pettinari and Chuvieco, 2016), but with a coarse spatial resolution, being uncapable for use at operational purposes that require a finer scale of analysis and accuracy. Aragoneses et al. (2024) provided a pan-European raster for the canopy base height (CBH) at 1 km as target resolution. Information on canopy bulk density (CBD) is difficult to estimate due the complexity of extracting the foliage biomass from the satellite sensors (see Table 1). The same team that generated the CBH model, produced a surface fuel map at the same spatial resolution (1 km) (Aragoneses et al. 2023).

The aim of this work is to provide a coherent set of raster files (Table 1) to better characterise the fire hazard, behaviour, and risk based on assessment analyses at a pan-European level, organized at different administrative divisions according to the nomenclature of territorial units for statistics - NUTS. The second goal was to share such data as per F.A.I.R. guiding principles (Fecher et al. 2015) for research data stewardship (Findability, Accessibility, Interoperability, and Reusability). These raster files are the necessary inputs needed to run many forest fire simulation methods. They are provided as open data available at an harmonized spatial resolution of 100 m to the public. End-users can access a user-friendly interface for downloading specific areas of interest autonomously, or the entire pan-European datasets.

## 2 Datasets provided on the fuel map open server

Current operational models that estimate fire behaviour require geospatial data in grid format (Cardil et al. 2023). This data is usually a set of N raster layers that are commonly used to model fire behaviour in any study area of interest. Some variables that we provide in this server were already generated and provided in other geo-portal servers by other authors in raster format at different spatial resolutions (see Table 1). However, here, we share the estimations of the most challenging input variables to be produced at the pan-European level: two canopy fuel layers (CBH and CBD) and the surface fuel model layer (Kutchartt et al. under revision).

Table 1 describes the raster datasets that are available in the Web-GIS portal that were resampled at approximately 100 m. The original spatial resolution of the topography variables and canopy cover was 30 m, while the canopy height was at 10 m. The aboveground biomass, the canopy fuels, and the fuel model were generated by the authors at 100 m.

**Table 1**. Description of the inputs of required geospatial data for running forest fire simulations. They are co-registered in the same reference system and cell resolution.

| Variable | Description raster file |
|---|---|
| | |

| Topography | Topographical information regarding **elevation**, **slope**, and **aspect** was derived by a global digital surface model (DSM) using the Panchromatic Remote-sensing Instrument for Stereo Mapping, which was onboard the Advanced Land Observing Satellite launched by the Japan Aerospace Exploration Agency, see Tanodo et al. (2016). |
|---|---|
| Biomass | **Aboveground biomass** (AGB; Mg/ha) data was derived using an ensemble of machine learning algorithms. The algorithms were trained on 49 covariates derived from remote sensing products, employing a stratified sampling approach. The training was based on a biomass map initially developed by the European Space Agency (ESA) in 2018, but the biomass map was predicted and updated to 2020 by Pirotti et al. (2023) with its own methodology in order to keep updating this input for the next years. |
| Canopy features | **Tree canopy cover** refers to the proportion (%) of the ground surface area obstructed by the vertical projection of tree crowns and is commonly obtained through satellite data (Hansen et al. 2013).<br>**Tree canopy height** is the vertical distance from the ground to the top of the trees (m), which was developed using a Deep Learning (DL) model combining GEDI and Sentinel-2 data (Lang et al. 2023).<br>**Canopy base height** refers to the vertical distance from the ground to the lowest continuous layer of live crown fuel in meters.<br>**Canopy bulk density** is the ratio of foliage biomass to canopy volume expressed in kg/m$^3$.<br>Both raster datasets were developed using remote sensing-derived products, artificial intelligence, and species-specific allometric equations (Kutchartt et al. under review). |
| Fuel model | The surface **fuel model** is a categorization/codification/classification of fuel typologies originally developed by Scott and Burgan (2005) to predict the potential fire behavior characteristics, which organize fire data geo-spatially by the vegetational structure, height, and moisture content. |

In the presented open-data server, we integrate the geospatial data required to support fire modelling. This data aims to characterise the terrain topography (point 1 in Table 1), and the vegetation that may be consumed by a fire (points 2-4 in Table 1). Overall, all 11 layers (3 topographic, 6 vegetation/fuel and 2 uncertainty) were loaded at the four European NUTS levels suggested by the EUROSTAT (2022) at the pan-European scale (Figure 1).

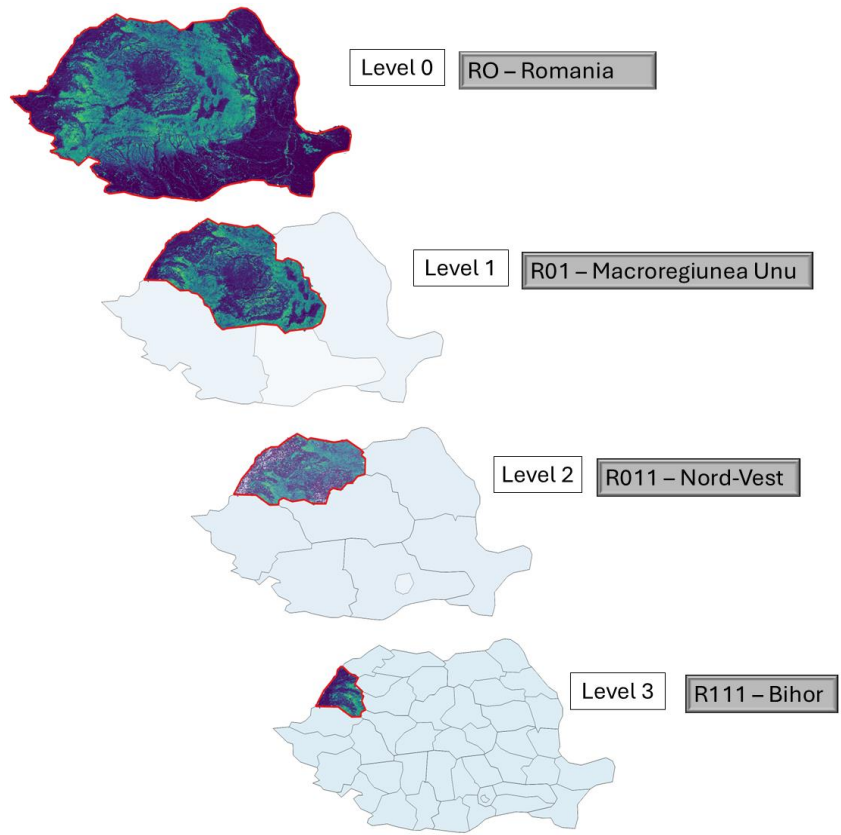

**Figure 1**. A demonstration of the layer levels available at the pan-European fuel server at all the NUTS levels from 0 to 3, for the case of Romania.

*2.1 Canopy feature rasters layers*

All raster layers were co-registered in the WGS84 geographic coordinate reference system (CRS) defined in the EPSG database version 10.008 with EPSG code 4326 with approximately 100 m cell size (8.983° x $10^{-4}$). It should be noted that this implies that cells will have different sizes according to the latitude of the cell, but the values in the cell node refer to a 100 m x 100 m area. We will therefore refer to pixel ground sampling distance as 100 m.

As indicated in Table 1, most of the layers in the geospatial data for fire simulators exist and were simply resampled at ~100 m and co-registered in the common reference frame and extent. In the framework of the H2020 "FIRE-RES" project, three specific layers were produced and delivered: CBH, CBD and surface fuel models. Also, the relative uncertainty maps for CBH and CBD were calculated and provided in the server. The pipeline to produce these layers starts with the estimation of above ground biomass (AGB) using artificial intelligence and a set of predictors that include bioclimatic variables, vegetation indices from Sentinel-2, and radar backscatter from Sentinel-1 and ALOS. A detailed explanation of the method is provided in Pirotti et al. (2023). Novel aspects regarding biomass estimation are found in segmenting the European area in several tiles to train independently with stratified samples to cover in a balanced way the range of biomass variability available in each tile. An ensemble of machine learners was thus trained and used for mapping AGB over Europe for the reference year 2020.

The aboveground biomass (AGB) map was then used for extracting the canopy bulk density (CBD), which is the amount of thin biomass in the trees - i.e., highly flammable fuel. Bulk density refers to the amount of thin biomass per unit volume of the canopy (Fig. 2). Thin biomass is mostly leafing biomass, thus AGB without trunk and branch biomass (Fig. 2). To estimate the thin biomass, allometric models were used to infer the fraction of thin biomass from the total AGB. The value of the fraction is species-specific and depends on the size and age of the tree. To account for the former, we used a map of tree species probability (Bonannella et al. 2022), while tree size and age were estimated by using the canopy height map by Lang et al. (2023) and inverse allometric models to derive diameter. Therefore, a set of biomass equations at tree components provided the fraction of thin biomass (foliage) for different tree species based on diameters as an explanatory-variable (Forrester et al. 2017).

Once the biomass of the thin fuel components is determined, it is possible to determine the density using canopy volume as the denominator. Canopy volume was determined using tree height and the estimated canopy base height (CBH). Subtracting CBH from total tree height provided the canopy volume (Fig. 2). Therefore, canopy fuels are fundamental variables considering that fire usually starts from the ground and increases in intensity and spread rate if it can reach the crown easily, especially when it is near the ground (understory fuels). A detailed explanation of the procedure for extracting CBH and CBD is available in Kutchartt et al. (under review).

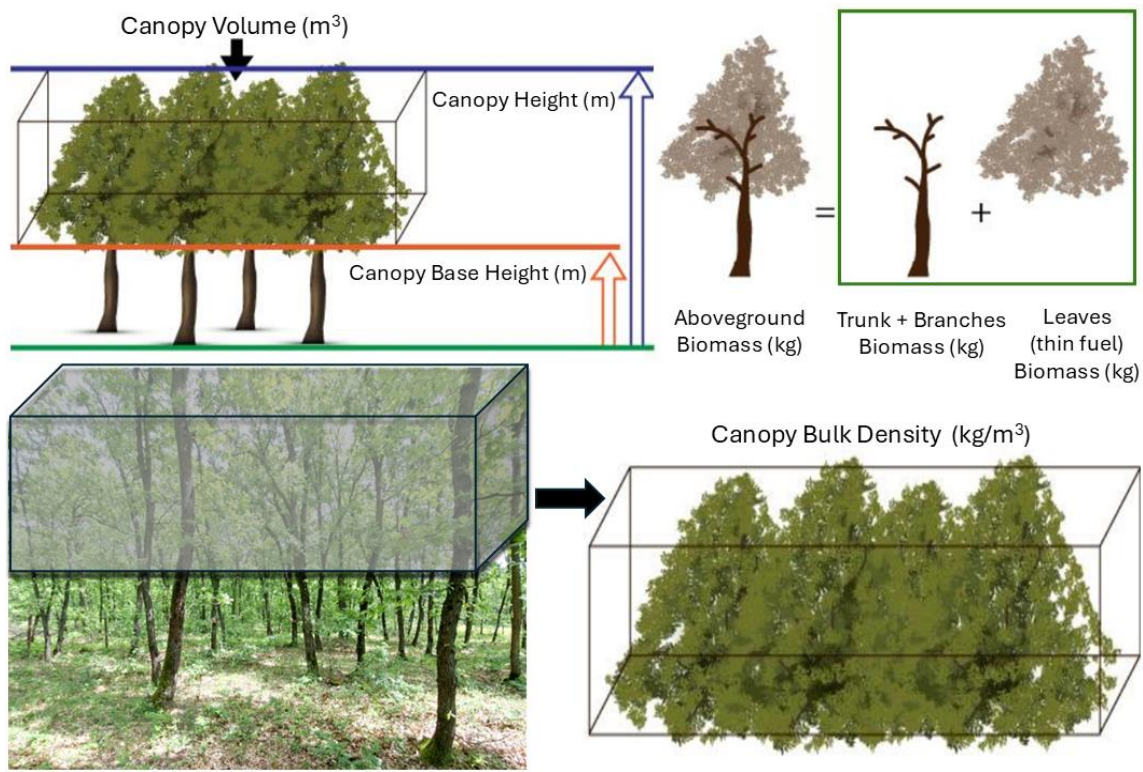

**Figure 2**. Clockwise from top left: schema depicting the relation between canopy volume, canopy base height and canopy height; Top right: schema of AGB biomass with respect to trunk, branches, and thin biomass; Bottom right: depicts only the foliage biomass component; Bottom left: is a geotagged image captured for validation to show how CBD is interpreted in the field.

*2.2 Fuel model raster layer*

The fuel model categories by Scott and Burgan (2005) divide fuels in three main burnable categories i.e., timber, shrub and grasslands, plus some non-burnable categories like urban, water, snow, and agricultural. Fuel model classes were determined according to the estimation of available fuel load and climate conditions (i.e. humid vs. dry environments). The fuel models were assessed and assigned to each cell through a consensus among different land cover maps to define the main categories at first. As in Aragoneses et al. (2023), several land cover maps are used together. The method differs in the number of land cover maps used and in the final attribution of categories. The following available datasets were used in the process: the CORINE 2018 at 100 m resolution, the Copernicus global land cover (GLC) at 100 m resolution, and the ESA World Cover 2021 v200 at 10 m resolution (based on Sentinel-1 and Sentinel-2 data).

Respectively, the overall thematic accuracies of these maps are >85%, ~80%, ~75%. The minimum mapping unit of the CORINE land cover is 25 ha. This means that usually features smaller than this size are clustered together, even if this is not a hard limit (Büttner et al. 2021). The canopy height map, canopy cover map and the AGB map were used for assigning the final fuel model category through a decision tree process. This procedure uses the canopy height to cross-checks that timber, shrub and grassland main categories have consensus (*e.g.*, forests have canopy heights > 2 m). It further uses canopy cover to provide a weight to cells that might have multiple land cover categories. The method provides a stack of rasters, one for each category in Aragoneses et al. (2023) with respective weights. A final map with the category is provided by selecting the category with the maximum weight from the stack for each cell. Further cross validation regarding the presence/absence of fuel material, and the estimate of the amount of fuel material was done also using the biomass map. The final Scott and Burgan (2005) categories were assigned after calculating an aridity index (AI) map estimated from bioclimatic data (precipitation, minimum average and maximum temperature, and solar irradiation) at 1 km spatial resolution from Hijmans et al. (2005). Climatic rasters were used to estimate evapotranspiration using the Hargreaves/Samani equation (Hargreaves and Samani, 1985) and final aridity from dividing precipitation by evapotranspiration. Six classes of AI (hyper-arid, arid, semi-arid, dry sub-humid, humid, and hyper-humid) as per FAO-UNEP classification were assigned to each cell. A final fuel model category was assigned to dry (arid to semiarid climate - rainfall deficient in summer) or humid (subhumid to humid climate - rainfall adequate in all seasons) subcategories as per Scott and Burgan (2005) according to the average AI value during the summer months. The final fuel model map was tested, and feedback was provided from living labs and in particular from experts involved in the project that used the data as input for fire behaviour simulations.

*2.3 Uncertainty layers of CBH and CBD maps*

The procedures described above include multiple estimation steps through modelling, which naturally will propagate the uncertainty of the original data and of the estimations along the pipeline of the workflow. It was therefore important to map an uncertainty layer for each final product and thus provide a further informative layer to the users. Two different approaches were used to carry out uncertainty analysis: the chain rule and Monte Carlo simulations. The former was used when the equations were relatively simple and were expected to have normally distributed values, such as the CBH map. The allometric model is a linear regression model so the expected error from the independent variable can be tracked. When functions become more complex like in the case of CBD, or require multiple steps like the fuel model map, then the Monte Carlo method is a more suitable tool. This method uses random samples of the input variables of interest extracted from an expected frequency distribution, described with an average and standard deviation, thus simulating the expected error.

*2.4 Pan-European fuel map server*

Google Earth Engine (GEE) provided the big-data infrastructure to organize all raster layers used and created in the project and also provided the means to share them through an app that was published online as the pan-European fuel map server (Fig. 3). The GEE environment allows linking data layers, loaded as GEE assets, to the web-based portal thus providing the necessary user-friendly interface for interacting with the data layers. R language environment was also used for some processing steps, such as training and applying the AI framework for estimating biomass for 2020 and for sub-setting the grid data at pan-European scale to the NUTS levels up to level 3. At the end of this last process, more than 1800 raster files are generated and stored for user download. The pan-European fuel maps server allows users to easily find and access each single layer of geospatial data. Visualization is provided as a styled raster layer. An interactive user query can extract all information from the raster layers at a user-defined location by selecting a location it with the left mouse button. Further interaction consists in downloading the layers or the whole stack at the user-selected area at the chosen NUTS level. This increases the accessibility of the data, as analysis and decisions are usually made at specific administrative levels, with respective boundaries, such as states, provinces or regions. The replicability of the process is assured through the documented methodology, and both the R code and the GEE code for the server are available upon request to the authors.

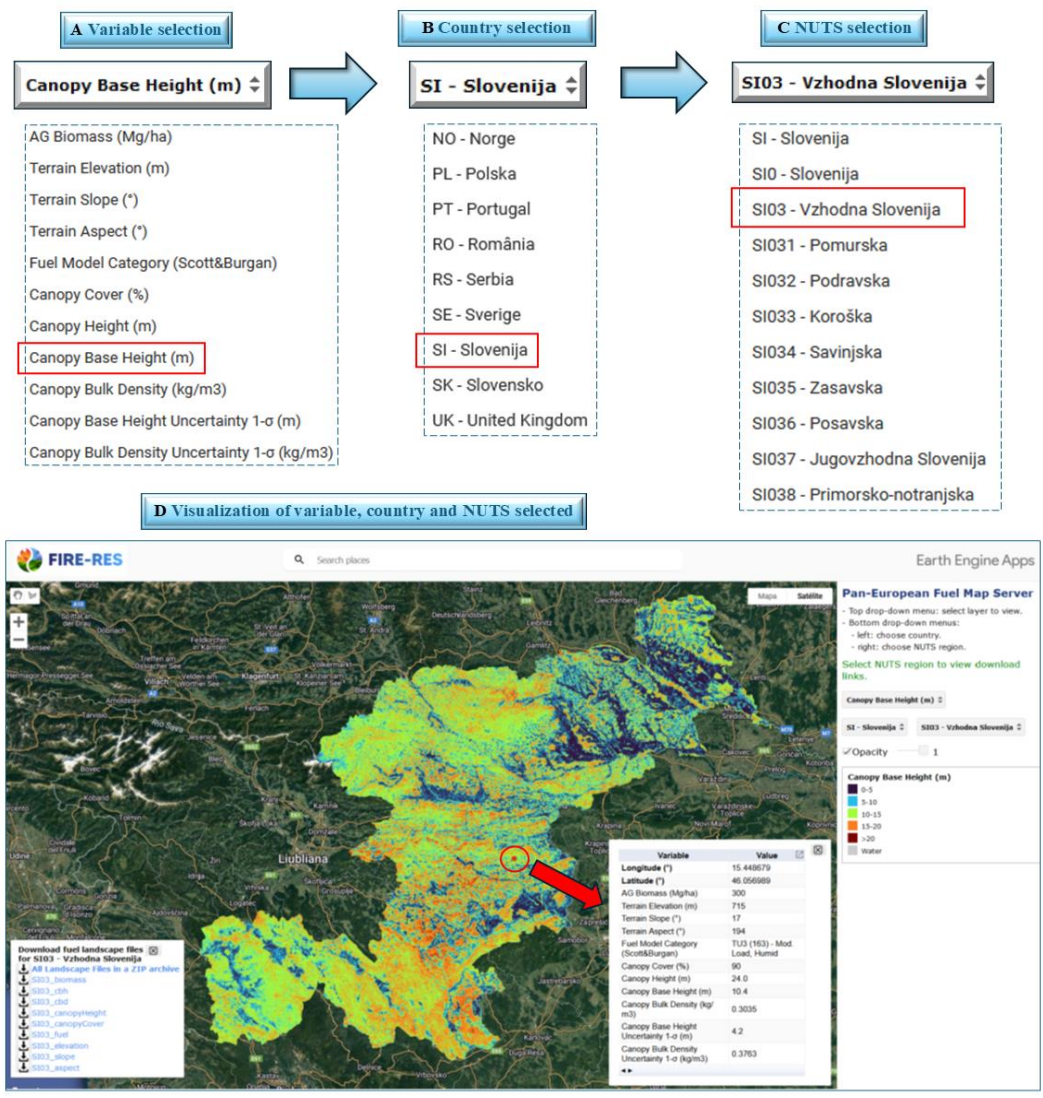

**Figure 3**. An overview of the workflow on how to download the raster data: a) selecting the variable, b) selecting the country, and c) the NUTS level. Once the variable and geographical area is defined, d) a visualization will appear with the variable, country, and NUTS chosen, where a box appears with the option to download all raster files in a zipped folder or single raster files is provided. In addition, the raster file can be queried at any pixel, once the user clicks on it, a red dot will appear and a table will be opened with the values of the eleven variables offered, including the nine raster layers available, and the two uncertainty layers reported for the CBH and CBD.

*2.5 Ground validation*

The pan-European fuel maps server allows users to view and interact with the data. An important part of creating and sharing geodata is the accuracy of the information. In this case, the challenge is the large area considered, and the variability of data inside the 100 m cell. To address this, a solution that uses crowd-sourced imagery is under development for improving the accuracy of the maps in the server. A smartphone application was developed in parallel with the pan-European fuel mapping server to support validation as follow-up research (Kutchartt et al. 2023). This app, called the FIRE-RES Geo-Catch app, differs from other photo-collecting apps as it is based on a very simple interface and uses a Progressive Web App (PWA) framework and thus can be accessed via a web-browser and/or be quickly installed in user's smartphones. It collects user photos as images with a specific EXIF tag enriched with geo-location, along with its accuracy, and camera orientation at the time the picture is taken. Thus, for a specific location, one or more images of the landscape with the direction of the line-of-sight becomes available in real-time after the user takes the picture. This supports validation at the corresponding cell at that location in the pan-European fuel maps server. In Kutchartt et al. (2023) Figure 2 shows a depiction of a geotagged and oriented image captured by the FIRE-RES Geo-Catch app and Figure 5 shows future development that will label images with categories of fuel models and other information. This step will be followed by training an AI model that will automatically predict a fuel model category for all the images. To this date 5000+ images have been collected throughout Europe.

## 3. Conclusions

This technical note reports on the role of geomatics for reaching two main goals, i.e., the creation of mapped products for decision support related to fire-risk management and sharing these data through a pan-European fuel maps server as per the F.A.I.R. guiding principles for research data stewardship. The geospatial data for fire simulations is used to refer to the co-registered set of gridded data that provides stakeholders key information to define fire prevention plans and other data-driven decisions as it is used by many fire behaviour models. Academia, public administrations, agencies and common citizens can find, query, and access the data at various scales to make informed decisions. Further research will focus on downscaling the fuel model and canopy fuel rasters at an even more operational scale, likely 30 m or less of ground sampling distance along with applying validation methods using LiDAR data and forest inventory plots as ground truth values.

**Funding**

This work was funded by the European Union's Horizon 2020 Research and Innovation Programme through the project entitled "Innovation technologies & socio-ecological-economic solutions for fire resilient territories in Europe - FIRE-RES" under grant agreement Nº101037419. In addition, the researchers Erico Kutchartt, José Ramón González-Olabarria, and Jordi Garcia-Gonzalo were supported through a mobility program by the project entitled "Decision support for the supply of ecosystem services under global change - DecisionES" under grant agreement Nº101007950. Dr. Jordi Garcia-Gonzalo and Dr. José Ramón González-Olabarria received funding from the Agency for the Research Centre of Catalonia (CERCA Programme/Generalitat de Catalunya). Dr. Erico Kutchartt was supported by Fondazione Cassa di Risparmio di Padova e Rovigo (CARIPARO). Dr. Francesco Pirotti was supported within the Agritech National Research Center and received funding from the European Union Next-GenerationEU (PIANO NAZIONALE DI RIPRESA E RESILIENZA (PNRR) – MISSIONE 4 COMPONENTE 2, INVESTIMENTO 1.4 – D.D. 1032 17/06/2022, CN00000022). Dra. Núria Aquilué was supported by a Juan de la Cierva fellowship of the Spanish Ministry of Science and Innovation (FCJ2020-046387-I).

**CRediT authorship contribution statement**

**Erico Kutchartt:** Conceptualization, Investigation, Formal analysis, Data curation, Methodology, Writing - original draft. **José Ramón González-Olabarria:** Conceptualization, Methodology, Supervision. **Núria Aquilué:** Conceptualization, Writing - review & editing. **Jordi Garcia-Gonzalo:** Conceptualization, Resources. **Antoni Trasobares:** Funding acquisition, Supervision, Resources, Project administration. **Brigite Botequim:** Conceptualization. **Marius Hauglin:** Data curation, Validation. **Palaiologos Palaiologou:** Conceptualization, Data curation, Validation. **Vassil Vassilev:** Conceptualization, Validation. **Adrian Cardil:** Conceptualization, Writing - review & editing. **Miguel Ángel Navarrete:** Data curation, Validation. **Christophe Orazio:** Conceptualization. **Francesco Pirotti:** Methodology, Data curation, Formal analysis, Software, Validation, Writing - original draft.

**Data availability**

All the data are available in the pan-European fuel maps server, where the nine raster files can be visualised, queried, and downloaded at different NUTS levels for the whole pan-European region. In addition, the uncertainties of the canopy base height (CBH) and the canopy bulk density (CBD) maps are also available. The open server was implemented via Google Earth Engine app in the following link: www.cirgeo.unipd.it/fire-res/app

**Declaration of competing interest**

Authors declare no known competing interest or personal relationship that could influence the integrity of this work.

**Acknowledgements**

The key contribution from FIRE-RES partners through the ten European living labs is acknowledged. They actively participated in the testing phase of our raster files during their forest fire simulations at different geographical locations that helped to improve the final products.